# Spectrally dispersed kernel phase interferometry with SCExAO/CHARIS: proof of concept and calibration strategies

**Alexander Chaushev,a,* Steph Sallum,a Julien Lozi,b Frantz Martinache,c Jeffrey Chilcote,d Tyler Groff,e Olivier Guyon,b,f,g Jeremy Kasdin,h Barnaby Norris,i and Andy Skemerj**

aUniversity of California Irvine, Department of Physics and Astronomy, Irvine, California, United States
bNational Astronomical Observatory of Japan, Subaru Telescope, Hilo, Hawaii, United States
cObservatoire de la Côte d'Azur, Laboratoire Lagrange, Nice, France
dUniversity of Notre Dame, Department of Physics, Notre Dame, Indiana, United States
eNASA-Goddard Space Flight Center, Greenbelt, Maryland, United States
fNational Institutes of Natural Sciences, Astrobiology Center, Mitaka, Japan
gUniversity of Arizona, Steward Observatory, Tucson, Arizona, United States
hPrinceton University, Department of Mechanical and Aerospace Engineering, Princeton, New Jersey, United States
iUniversity of Sydney, Sydney Institute for Astronomy, School of Physics, Sydney, New South Wales, Australia
jUniversity of California, Santa Cruz, Department of Astronomy and Astrophysics, Santa Cruz, California, United States

**Abstract.** Kernel phase interferometry (KPI) is a data processing technique that allows for the detection of asymmetries (such as companions or disks) in high-Strehl images, close to and within the classical diffraction limit. We show that KPI can successfully be applied to hyperspectral image cubes generated from integral field spectrographs (IFSs). We demonstrate this technique of spectrally dispersed kernel phase by recovering a known binary with the SCExAO/CHARIS IFS in high-resolution K-band mode. We also explore a spectral differential imaging (SDI) calibration strategy that takes advantage of the information available in images from multiple wavelength bins. Such calibrations have the potential to mitigate high-order, residual systematic kernel phase errors, which currently limit the achievable contrast of KPI. The SDI calibration presented is applicable to searches for line emission or sharp absorption features and is a promising avenue toward achieving photon-noise-limited kernel phase observations. The high angular resolution and spectral coverage provided by dispersed kernel phase offers opportunities for science observations that would have been challenging to achieve otherwise. © *The Authors. Published by SPIE under a Creative Commons Attribution 4.0 International License. Distribution or reproduction of this work in whole or in part requires full attribution of the original publication, including its DOI.* [DOI: 10.1117/1.JATIS.9.2.028004]

**Keywords:** kernel phase; high angular resolution; calibration; data processing.

## 1 Introduction

Kernel phase interferometry (KPI) is a data analysis method that can resolve angular separations at or below $\lambda/D$ by treating a conventional telescope as an interferometric array.[1] KPI probes similar angular separations as related techniques, such as nonredundant masking,[2] but without the corresponding loss of throughput that comes from using a pupil mask.[3] By producing interferometric observables that are self-calibrating to first order in phase, KPI can be used to identify close-in asymmetric signals, such as companions or skewed circumstellar disks, that may be inaccessible with conventional imaging.

---
*Address all correspondence to Alexander Chaushev, a.chaushev@uci.edu

Given the large distances to young stars,[4] high angular resolution, spectrally dispersed observations are key to solving open problems in planet formation and circumstellar disk evolution.[5] This strongly motivates the application of KPI on new instruments. Its high throughput means that KPI is amenable to not just discovery but also characterization. For example, applied on high-Strehl imaging spectrographs (e.g., behind extreme adaptive optics systems and in space), KPI could enable effcient, spectrally dispersed observations where building signal to noise with a pupil plane mask would be time prohibitive.

In this paper, we study the application of KPI to the Coronographic High Angular Resolution Imaging Spectrograph (CHARIS),[6-8] an integral field spectrograph (IFS) for the Subaru telescope. As an IFS, CHARIS generates a spectrum for each point in its field of view, providing a hyperspectral cube of images at a range of wavelengths from which kernel phases can be calculated. CHARIS also sits behind the Subaru Coronagraphic Extreme Adaptive Optics (SCExAO) system,[9-11] which is critical since KPI requires that the incoming wavefront is well corrected (residual phase errors are small) with a Strehl ratio of ∼80% or greater.[3,12] When observing on the ground, this is only possible behind a powerful adaptive optics system. At present, the only ground-based IFSs capable of meeting these Strehl requirements are CHARIS, GPI,[13] and SPHERE.[14] To-date, KPI has not been demonstrated on any of these instruments. Demonstrating spectrally dispersed KPI would be a first, enabling a range of new and more sensitive interferometric observations.

Currently, many KPI observations are limited by high-order, residual systematic noise in the kernel phases.[12] Conventionally, for broadband KPI, these systematics are corrected by observing an unresolved point-spread function (PSF) reference star, which is then used as a calibrator. However, this type of reference differential imaging (RDI)[15] calibration only accounts for common systematics between the science target and PSF calibrator. Typically, a reference calibrator will have a different spectral type, magnitude, and airmass from the science target, as finding an identical calibrator close on-sky is challenging. This can result in uncorrected "calibration errors," since the PSF KPs do not accurately represent the systematics present in the science target.[12]

A key goal of this study is to compare the performance of kernel phase RDI (KP-RDI) and kernel phase spectral differential imaging[16] (KP-SDI) for calibration of data. The latter uses the wavelength information provided by the IFS to calibrate residual systematics. For CHARIS, due to the relatively small width of the spectral bands (∼20 nm) and simultaneous measurement, kernel phases from adjacent wavelength bands may provide a better calibration than dedicated PSF stars. In this case, the data would be "self-calibrating" even beyond KPI's first-order elimination of instrumental phases, requiring no additional observations of other targets. This would be advantageous since more telescope time could be devoted to the science target, maximizing the signal-to-noise ratio ($S/N$) and increasing observational efficiency.

In addition to assessing the viability of KPI with CHARIS and KP-SDI, we discuss the calibration quality within the context of a science case where KPI is used to search for excess Br$\gamma$ emission from a protoplanet orbiting a young star. Br$\gamma$ emission would signal that the protoplanet is still accreting material[17] and could elicit information about the detailed physics of planetary formation. For this scenario, one would search for an excess signal above the continuum in a single spectral bin, making SDI calibrations relatively simple.

## 2 Kernel Phase Theory

KPI treats a conventional telescope as an interferometric array by discretizing the pupil into a set of virtual subapertures. We can build a model between instrumental pupil-plane phases ($\phi$) and the phases measured from the Fourier-transformed images ($\Phi$):

$$\Phi(u, v) = \arg \Sigma_{ij} \exp^{i(\phi_j - \phi_i)},$$ (1)

where the indices $i$ and $j$ represent the pairs of subapertures in the pupil model sampling spatial frequencies ($u, v$). At high Strehl (i.e., low residual phase errors), we can Taylor expand the exponential terms for each spatial frequency, recovering a linear relationship between the pupil plane phases and Fourier phases.

We can then write the ensemble of Fourier phases as a vector:

$$\Phi = R^{-1} \cdot A \cdot \phi + \Phi_0, \tag{2}$$

where $\Phi_0$ represents the vector of science target phases measured at a particular wavelength, $R^{-1}$ encodes the redundancy of each spatial frequency sampled by the pupil model, and $A$ describes which pairs of subapertures contribute to each spatial frequency. The dot product $R^{-1} \cdot A$ thus describes how the pupil plane phases map to the phases measured in Fourier domain.

For an IFS, the values of $\phi$ and $\Phi_0$ will be dependant on the wavelength, in which the observations are conducted, with each wavelength corresponding to one distinct spectral band of the IFS' overall wavelength coverage. Considering each spectral band separately, using singular value decomposition we can then find a matrix $K$, such that $K \cdot A = 0$. This is the left nullspace, or kernel, of $A$. Left-multiplying Eq. (2) through by $K \cdot R$ gives

$$K \cdot R \cdot \Phi(\lambda) = K \cdot R \cdot R^{-1} \cdot A \cdot \phi(\lambda) + K \cdot R \cdot \Phi_0(\lambda), \tag{3}$$

$$= K \cdot A \cdot \phi(\lambda) + K \cdot R \cdot \Phi_0(\lambda), \tag{4}$$

$$= K \cdot R \cdot \Phi_0(\lambda), \tag{5}$$

where $K \cdot R \cdot \Phi_0$ are the "kernel phases" as a function of wavelength. $K$ and $R$ remain the same for different wavelengths as the pupil model is the same. In this way, first-order perturbations of the phase, due to residual AO errors, are eliminated, and the kernel phases can be used for modelling or image reconstruction to recover asymmetries present in the observed target.

Considering the wavelength dependence further, since the kernel phases are computed from images taken simultaneously at slightly different wavelengths, the same $(u, v)$ sampling point in adjacent spectral bands will sample phases [$\Phi$ in Eq. (3)], which correspond to slightly different spatial frequencies on the sky (measured in arc $\sec^{-1}$). This is due to the wavelength dependence in spatial frequency $b/\lambda$, where $b = \sqrt{u^2 + v^2}$. This means that for a science signal, where the source is at the same angular position relative to the PSF as a function of wavelength, the kernel phase signal will change from spectral band to spectral band. This is different from the behavior of (achromatic) higher-order systematic noise sources that are fixed relative to the PSF, since they make phase contributions to the same $(u, v)$ sampling points independent of wavelength. In practice, systematic errors are more complex and not necessarily achromatic, and this can be assessed by measuring the correlation in the kernel phases across wavelengths.

## 3 Observations

We conducted observations over the course of a quarter night on March 19, 2021, with SCExAO/CHARIS[6–8] on the Subaru Telescope. CHARIS is an IFS with resolving power $R = 70$ to 90 when observing in J, H, or K-band high-resolution mode. CHARIS produces images with a $2.07 \times 2.07$ arc sec field of view and a spatial scale of 16.2 mas per lenslet. We observed using the high-resolution K-band mode, which has an average resolving power of $R = 77.1$ and 17 wavelength bins covering 2.015 to 2.368 nm. Observations were conducted with a 10:90 beamsplitter, with 10% of the light going to CHARIS to avoid saturating on the targets, and the remaining light going to the wavefront sensor. Such bright targets were chosen to ensure sufficient light in the wavefront sensing band for optimal AO performance.

Table 1 has a summary of the observations, which include a known binary (HD 44927), a PSF reference star (HD 249005), and a telluric calibrator (HD 35036). The PSF reference star and telluric were chosen using the SearchCal tool.[18,19] SearchCal uses the empirical relationship between stellar angular diameters and photometries to estimate the stellar diameter of a star given a spectral type and VJHK magnitudes.[20] In addition to the frames listed in Table 1, dark and flat fields frames were taken at the start of the night for data calibration.

**Table 1** Observing log for March 19, 2021.

| Identifier | Type | Gaia $R_p$ | H-band | K-band | $t_{frame}$ (s) | $n_{frames}$ | Δ PA (deg) |
|---|---|---|---|---|---|---|---|
| HD 44927 | 50-mas binary | 6.0 | 5.9 | 5.9 | 20 | 53 | 2.1 |
| HD 245009 | PSF calibrator | 7.5 | 6.0 | 5.8 | 20 | 42 | 1.2 |
| HD 35036 | Telluric calibrator | 7.1 | 6.0 | 5.8 | 20 | 52 | 0.1 |

The seeing for the quarter night was ∼0.4 arc sec from data available from the Mauna Kea Weather Center. The Strehl ratio was estimated by comparing the PSF of the reduced K-band images to a simulated CHARIS PSF generated using a model of the pupil. It was found to be above 0.8 throughout the observations.

## 4 Data Processing

### 4.1 Reduction

The hyperspectral data cubes were constructed using the CHARIS-DEP reduction pipeline,[8] which first calibrates raw images and then extracts the spectra to form a cube of images at the various wavelength bins. First, the flat fields and dark frames were processed using the "buildcal" tool to produce master calibration frames. Next the "extractcube" command was run over the raw images. The spectral extraction was completed using the least-squares method, with the "fitbkgnd," "suppressrn," and refine parameters set to "true." These allow for a lenslet-by-lenslet background subtraction, suppress read noise, and remove cross-talk from neighboring pixels. Figure 1 shows examples of the extracted images, where each pixel represents the incident light onto one lenslet in CHARIS' lenslet array. The pixels are aligned with the lenslets, and their values are the result of the least-squares fit to the lenslet PSFs making use of both instrument calibrations and calibrations taken during the night of the observations. Since the lenslet array is rotated relative to the CHARIS detector, and light from all of the lenslets does not fall onto the detector, the signal in the extracted images lies a rotated, square region of the final datacube. The reduced data cubes were then manually inspected and a wide-separation point-like feature was found in images from band 10 (2230 nm; Fig. 1) and to a lesser level in the adjacent bands 9 and 11. This is a ghost due to a known internal reflection, and was found consistently in all images. During preprocessing of the data (Sec. 4.2), the images were subframed to exclude the ghost, removing any potential effect on the KPs. After manual inspection, all data cubes were included in the subsequent analysis.

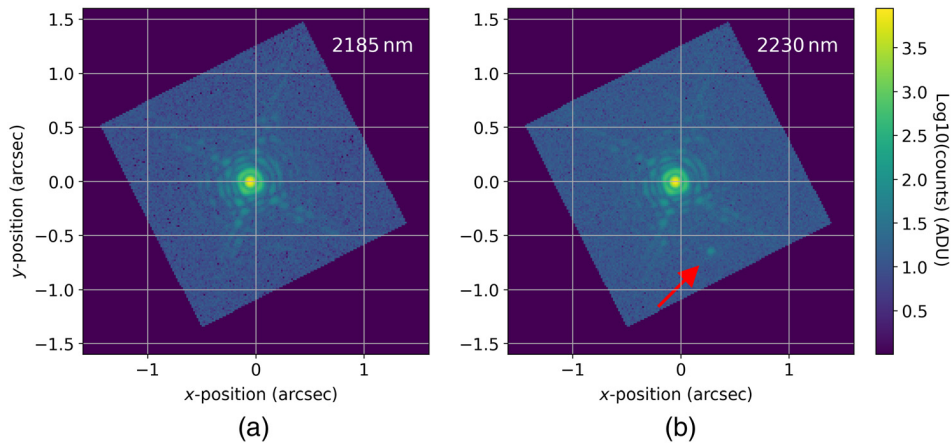

**Fig. 1** Two reduced CHARIS K-band images of HD 245009 (PSF calibrator) from the same data cube, showing bands 8 and 10 [(a) 2185 and (b) 2230 nm], respectively. The images have been background subtracted and log-scaled to highlight the structure of the PSF. The small spot between the diffraction spikes in the right-hand image (indicated by the red arrow) is a ghost which is due to internal reflections in the instrument. The feature is present in band 10 (2230 nm) for all targets and can also be faintly seen in the adjacent bands 9 and 11. During preprocessing, the CHARIS images are clipped down to a size of 64 × 64 pixels which excludes the artefact prior to the calculation of the kernel phases.

## 4.2 Image Preprocessing

We first subtracted the median background value for each extracted image to remove any residual background not removed by CHARIS-DEP. The images were then centered to the nearest pixel using an estimate of the image center computed using the XARA pipeline's "BCEN" centroiding algorithm.[1,21,22] The algorithm iteratively computes an image centroid after decreasing the size of a window around the PSF, until the centroid value converges within a given threshold. This ensures that the centroid is not affected by any bad pixels or other outliers.

A "nearest-pixel" centering approach is not sufficient for the purposes of KPI, since the technique is sensitive to low-level asymmetries in the PSF. A centroid shift adds a ramp in phase to the Fourier transform of the image, which can masquerade as a close-in asymmetry. For example, the expected phase signal for a close-in binary is a sinusoid in $(u, v)$ space, which would look nearly identical to a phase ramp at spatial frequencies that are low compared to the binary separation. Therefore, a second, subpixel centroid correction was computed by a grid search, in order to find the offset which minimized the standard deviation of the Fourier phases [computed across the unique $(u, v)$ sampling points]. The grid size was chosen to range from $-1$ to 1 pixel with an increment of 0.01 pixel. For each grid point, a phase ramp was subtracted corresponding to that centroid position and the standard deviation of the phases was computed. We subtract the phase ramp corresponding to the best centroid estimate from each image before calculating the kernel phases. Figure 2 shows an example of the phases before and after centroiding with the grid-search method.

We also use the standard deviation across kernel phase index to assess the quality of this centroiding approach, since lower levels of systematic error will result in a smaller spread of kernel phase values. Applying this centroiding procedure independently for each image in a spectrally dispersed CHARIS datacube produced kernel phases with lower standard deviations across kernel phase index than applying the same centroid to all images in the cube. The difference between the two centerings may be due to atmospheric dispersion causing the PSF to shift subtly between images of different wavelengths. Small differences in IFS extraction quality from wavelength bin to wavelength bin could also introduce subpixel shifts in the processed datacubes. Without centroiding on an individual basis, differences in the kernel phases between adjacent bands may be dominated by these chromatic effects, reducing the effectiveness of spectral differential calibration.

After centroiding, each image was cropped to a square size of $64 \times 64$ pixels. Several crop sizes were tested by computing the standard deviation of the kernel phases across kernel phase index for each image and then calculating the average value for all images in the dataset. Although a range of image sizes produced comparable KPs, 64 pixels resulted in the lowest mean standard deviation. Using this window, the ghost was also removed from the images, thereby having no effect on the kernel phases. The cropped images were also inspected to check that the distribution of the data close to and away from the ghost was the same. Finally, to remove any sharp discontinuities in the data and reduce any remaining random noise due to imperfect sky background subtraction, a super-Gaussian window function of order four and half-width half-max of 20 pixels was applied to each image.

## 4.3 Kernel Phase Calculation

We next computed the kernel phases from the preprocessed images. First, a model of the SCExAO pupil was rotated to match the CHARIS configuration. This rotation of $-69.5$ deg takes into account the angle between the SCExAO pupil and the CHARIS instrument, as well as the lenslet array rotation of $-63.5$ deg. Consistent with previous studies,[22] we used a "gray" model that so each grid point has an associated transmission factor, which can vary between zero and one (as opposed to a "binary" model containing only zeros and ones).

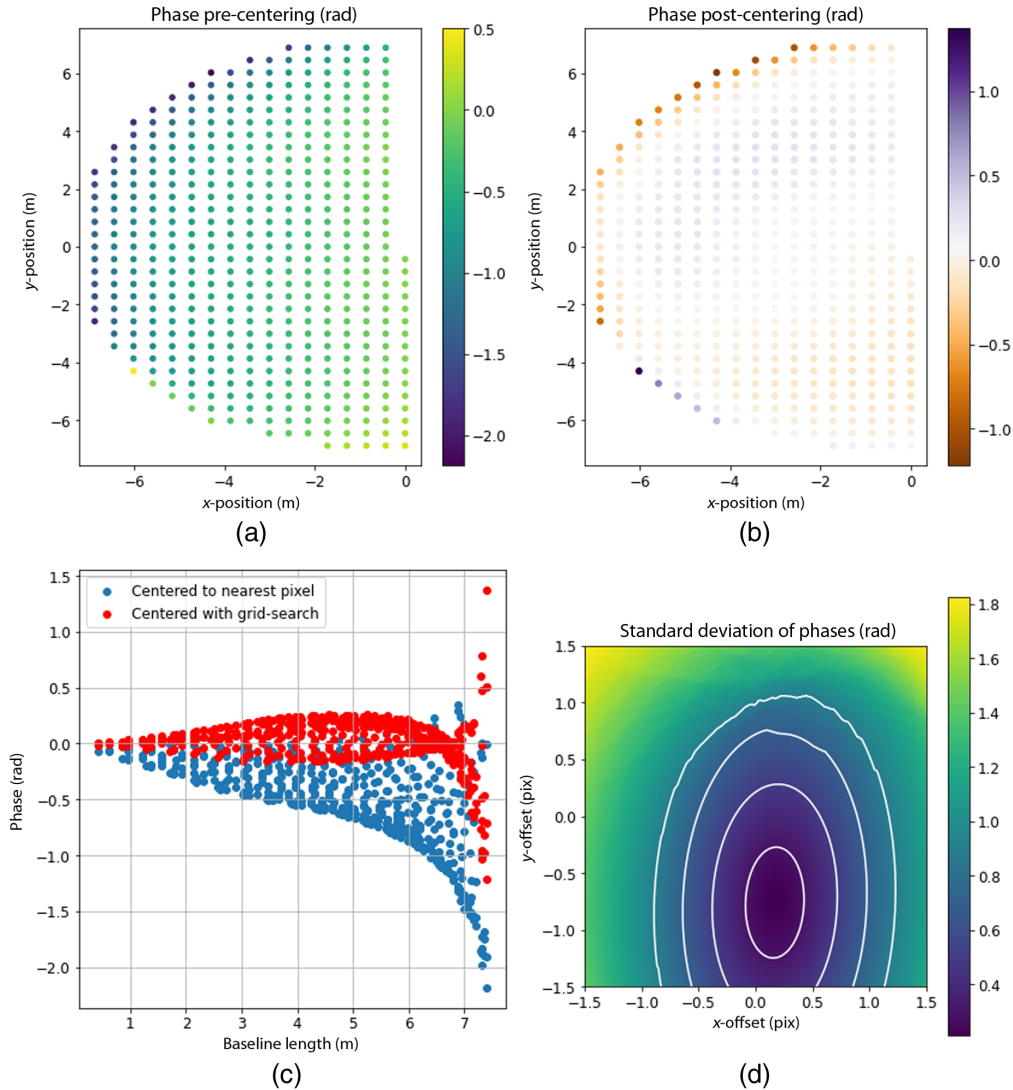

**Fig. 2** An example of the grid search subpixel centring process for an image from the PSF calibrator. (a) The measured phases for each UV-sampling point after an initial centering to the nearest pixel has been applied. (b) The phase after the fine recentring. (c) Phases before and after correction as a function of baseline length. (d) Standard deviation of the phases as a function of the x- and y-offsets explored during the grid search.

Next the pupil model was converted into a grid of discrete sampling points. The spacing of the pupil grid points sets the lowest spatial frequencies measured in the FT and thus the maximum spatial scale that the kernel phases can constrain. The considerations in setting the pupil model spacings are: (1) the size of the subframed images (64 pixels, corresponding to 1.036 arc sec) and (2) the width of the super-Gaussian window function [full-width at half maximum (FWHM) of 40 pixels, corresponding to 0.648 arc sec]. As described by Martinache et al.,[22] a baseline of length $b$ would probe spatial frequencies corresponding to fields of view (FOV) of $\lambda/b$, which sets a lower limit on $b$ of $b > \lambda/\text{FOV}$. This lower limit is equal to 0.43 m for the field of view of the cropped images. The super-Gaussian window also imposes a cut-off that (for its smaller FWHM) corresponds to a larger minimum baseline. However, following Martinache et al., we use the image crop size to set the minimum subaperture separation, since the super-Gaussian attenuates but does not fully eliminate signals on spatial scales greater than its FWHM.

With a lower limit of 0.43 m, we use injection and recovery tests for a range of model binaries to choose final pupil geometry parameters. In addition to selecting the grid spacing, we also

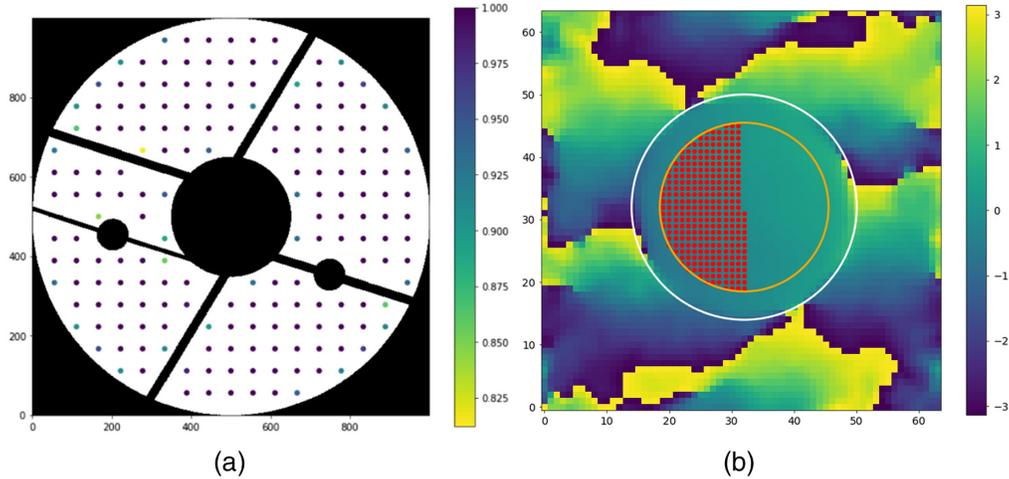

**Fig. 3** (a) The SCExAO pupil model used for the calculation of the kernel phases. The chosen subapertures are shown and each one is color-coded according to its transmission value. (b) The $(u, v)$ sampling points for the pupil model overlaid onto the phase values of the Fourier transform of an example image (at 2119 nm) from the PSF star data cube. The white-line marks the 7.74-m pupil diameter while the orange-line marks the effective 5.8-m cut-off, which is applied to remove noisy longer baselines from the $(u, v)$ plane.

varied the maximum baseline length by masking the highest spatial frequencies in the Fourier transform. This is justified since AO correction may be poorer on their corresponding fine angular scales and removing them from the phase vector may increase the signal to noise of the KPs.[3] For grid spacing, we tested a range of values from 1.26 down to 0.43 m, whereas for maximum baselines the range of parameters varied between 4 and 7.74 m. For each pair of parameters, we considered the standard deviation in the calibrated kernel phases across index (estimating the systematic noise level) and the standard deviation across kernel phase index of a range of model binaries (estimating the signal amplitude) to find the parameters which maximized this signal-to-noise ratio. One thousand model binaries were used for this testing, with parameters covering separations of 30 to 80 mas, contrasts from 0.5 to 4 magnitudes and position angles (PAs) spanning 0 to $2\pi$.

A 0.43-m spacing and 5.8-m $(u, v)$ cut-off maximizes the signal to noise, even in the case for binaries close to and below $\lambda/D$ (for SCExAO's pupil diameter of 7.74 m) where the longest baselines would provide the best constraints on the model binaries. Although this seems counter intuitive, examination of the PSF calibrator Fourier phases shows that the longest baselines have much larger standard deviations, which would inflate the kernel phase scatter and degrade the achievable contrast. The optimal cut-off varies by ∼1 m as a function of wavelength band. However, most show an increase in the Fourier phase error above ∼5.8 m. As shown in Fig. 3, the final pupil geometry, using a 0.43-m grid spacing, resulted in 182 subapertures in the full pupil model and 288 distinct UV-sampling points with baselines less than the 5.8-m cutoff.

The pupil model was then used to calculate the kernel phase matrix using the XARA pipeline and as described in Sec. 2. To calculate the kernel phases, for each subpixel-centered image, a discrete Fourier transform was taken at the $(u, v)$ sampling points corresponding to distinct baselines from the pupil model. Phases were then extracted by taking the angle of the resulting complex visibilities. Finally, a total of 107 kernel phases were computed from the measured phases for each image, by taking the dot product of the phases with the kernel phase and redundancy matrices as described in Sec. 2.

## 5 Validating CHARIS Kernel Phase with the Binary HD 44927

Figure 4 shows a histogram of the kernel phases for all three of the targets. Evidence that HD 44927 is a resolved binary can clearly be seen in Fig. 4, since the kernel phases show significantly higher spread in absolute value than those of the calibrator PSF and telluric. Prior lunar occultation observations of HD 44927 (from 1973) reveal a companion with a separation of 67 mas and PA 124 deg measured east of north.[23] We use KPI to recover this binary signal, expecting a similar, but not identical separation since the companion should have exhibited orbital motion between the lunar occultation and the data presented here.

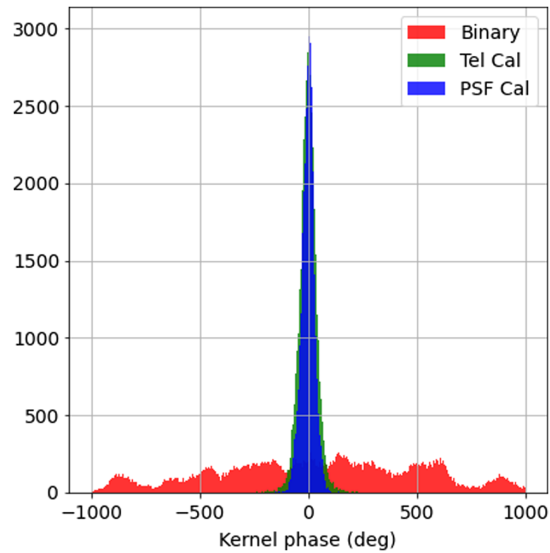

**Fig. 4** Histogram of the raw kernel phases, in degrees, for all three targets as computed from the XARA pipeline. The binary kernel phases have a flat non-Gaussian distribution, indicating a strong signal. The PSF and telluric calibrators have much smaller standard deviations (across kernel phase index) of approximately equal size, centered around a value of zero degrees.

To demonstrate CHARIS/SCExAO KPI, the observations of the binary at each spectral band were fit with a companion model in order to recover the separation, contrast, and PA of the system. Both the binary and the PSF calibrator datasets have very little parallactic angle evolution, of ∼2 deg and ∼1 deg, respectively, so the kernel phases from all frames were averaged for each target and band, respectively. The binary data were then calibrated using RDI, by subtracting the kernel phases of the PSF calibrator. The errors on the kernel phases were estimated by computing the sample covariance matrix of the kernel phase indices across all of the frames taken in each band. The diagonal elements were then taken, for both the binary and PSF calibrator, and added in quadrature to properly account for effect of the RDI calibration. The resulting errors are necessarily underestimates of the true errors since they do not take into account the residual systematics and the correlations between the different kernel phases. To quantify this difference we calculated the reduced chi-squared residual between the data and an MCMC model fit computed below. The value was 53.1 for the 19 parameter fit (of one separation, one PA, and 17 contrasts), indicating that the errors are indeed underestimated. To better account for these residual systematic errors and place more realistic uncertainties on the binary parameters, following Kraus and Ireland,[24] we inflated the errors by adding 0.35 radians to each one, such that the new reduced chi-squared value is ∼1.0.

Using the time-averaged kernel phases, the binary was fit in a two step process. First, a rough solution was found by comparing the kernel phases to a grid of single companion models. The grid contained 172,800 models with separations of 20 to 80 mas and step size of 2.5 mas, contrasts of 0 to 7 magnitudes with a step size of 0.07 mags and a PA step size of 5 deg. Setting the separation and PA to be constant across all spectral bands, and fitting each contrast individually, the best solution was a separation of 52.5 mas, a PA of 50 degrees and contrasts ranging from 0.7 to 0.9 magnitudes depending on wavelength. An independent grid fit to each spectral band resulted in very similar solutions with differences of at most 1 step size, between the best fit for the band and the global solution. Correcting for the rotation of the instrument, the true PA is 328 deg measured east of north.

Following the grid, the "scipy.optimize" package was used to refine the grid solution, using the default "BFGS" algorithm without any bounds on the binary parameters. Finally, an MCMC was run, using the "emcee" package,[25] with 64 walkers for 11,000 steps to further refine the solution and estimate the uncertainties. The MCMC solution gives a separation of $53.63 \pm 0.04$ mas, a PA of 325.05 deg $\pm 0.05$ deg and a contrast ranging from 0.75 to 0.90 magnitudes for the different wavelengths, respectively. The full corner plot for the fit can be seen in Appendix A. The posteriors are Gaussian and show no evidence of the contrast separation degeneracy which is present in kernel phase observations at small angular separations.[26] This is unexpected given that the binary separation is just below $\lambda/D$ considering the full telescope size of 7.74 m or at 0.7 lambda/D considering the maximum baseline length of 5.8 m (for the central wavelength). However, it may be explained by the high binary signal to noise due to its very low contrast and the excellent seeing and SCExAO performance on the night of the observations.

Figure 5 shows the model and observed kernel phases for four of the seventeen spectral bands. The near one-to-one correspondence indicates that the model explains the majority of the kernel phase signal. Both the separation of the binary and the PA agree with publicly available orbital parameters from the Washington Double Star catalogue.[23,27] This shows that a consistent physical solution for the binary has been found and that KPI can be successfully applied to CHARIS data.

Figure 6 shows the spectrum of HD 44927 B in units of contrast with respect to HD 44927 A, as well as expectations for the contrast ratio if the secondary is of either an A2 or A3 spectral type. Excluding the final band, the average contrast error is ~0.01 magnitudes, whereas the standard deviation of the contrasts as a function of wavelength is 0.024 magnitudes. This indicates that there may be some residual systematic noise affecting the binary fit and that the errors are not taking this into account sufficiently. Similarly, Fig. 7 shows the residuals between the observed and model KPs as compared to the uncalibrated PSF and telluric calibrators. The residuals have a larger range of absolute values than those of the calibrators. This may indicate that the RDI calibration has not removed all of the systematics in the data. As discussed previously, these systematics could come from temporal changes in PSF quality between the binary and calibrator, differences in spectral type, or differences in airmass. Both cases point to the fact that the errors on the kernel phases are likely underestimated, even after inflating such that the reduced chi-square value is equal to 1.0. This approach likely does not fully reflect the true size of the errors, since inflating by a single value is an oversimplified way of dealing with residual systematics because it neglects the covariance of the kernel phases.

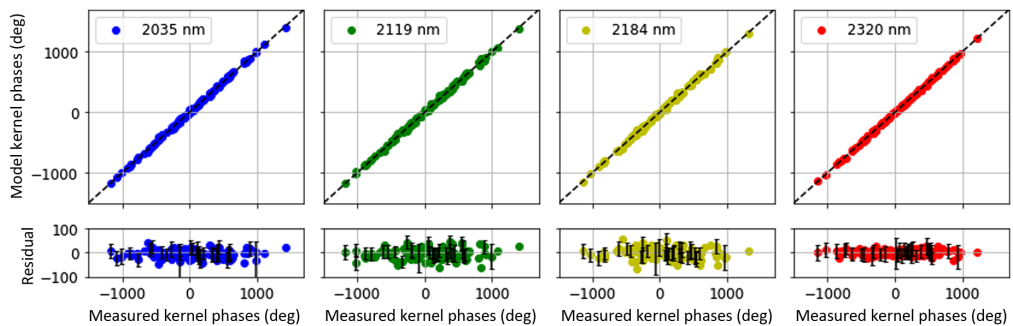

**Fig. 5** Comparison between the measured kernel phases ($x$ axis) for the known low-contrast binary HD 44927 and those of the binary model fit ($y$ axis). The 1:1 correspondence (dashed black) line can be seen in each case. The different colors show the fit for different bands, with the wavelength indicated in the figure legends. The kernel phases closely follow the 1:1 correspondence line, indicating that the model fits the data very well. Residuals betwen the model and the data can be seen in the bottom panels. Error bars have been placed on every third data point of the residuals. The errors have been inflated such that the reduced chi-squared fit of the best-fit model is ≈1.0 (see Sec. 5). The separation, PA, and contrasts extracted from the binary are consistent with literature values.

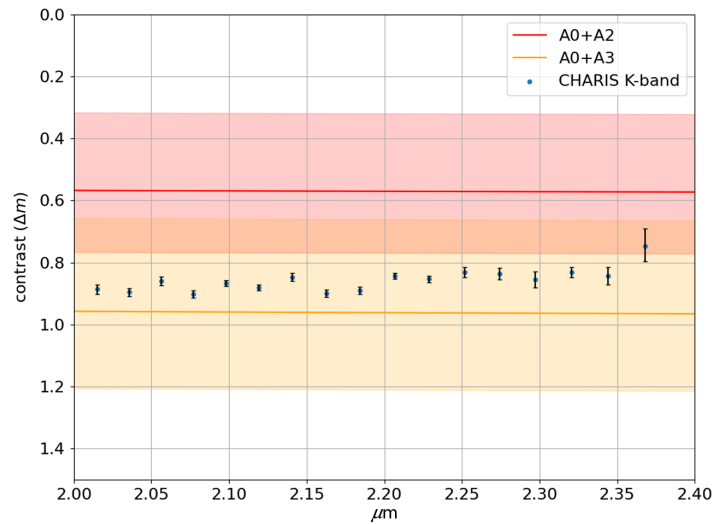

**Fig. 6** The spectrum of HD 44927 AB (blue scattered points) as computed from the high-res K-band CHARIS observations using an MCMC fit to the RDI-calibrated kernel phases. The red and orange regions denote the expected contrast if HD 44927 AB consists of an A0 primary with an A2 or A3 secondary, respectively. Error bars (black) have been added to each data point. The error bars have been inflated such that the reduced chi-squared of the best-fit model is equal to one. The contrast variation is large compared to the estimated error bars seen in the K-band spectrum and is likely due to the model fitting residual systematic variations. Although the errors have been inflated using the reduced chi-squared fit to the data, this method is not a perfect estimator of how underestimated the errors are.

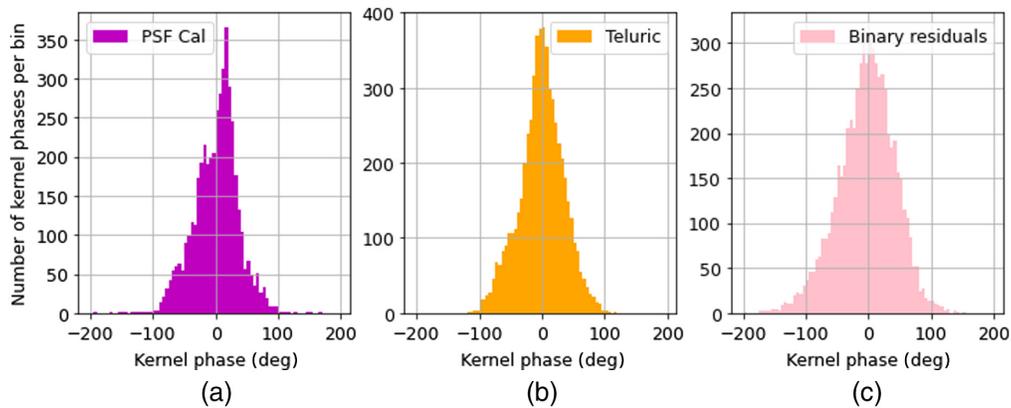

**Fig. 7** A comparison of the kernel phases of (a) the PSF calibrator (magenta) and (b) telluric (orange), and (c) the RDI calibrated binary data after the best fitting model of the companion has been removed (pink). All kernel phases are included in this figure, regardless of the wavelength band. The residual phases for the binary are slightly higher than the PSF and telluric KPs, indicating that there may be some residual systematic error that has not been correctly calibrated.

## 6 Comparing RDI to SDI with Kernel Phase

A typical RDI calibration strategy involves subtracting the kernel phases of the reference star from those of the science target. Here we compare this to a second scenario, an SDI calibration, where the "reference star" is constructed from the kernel phases of the science target which were taken simultaneously, but at adjacent wavelengths. Using SDI in this way, any continuum emission that is consistent across adjacent bands would be removed via the calibration. This may be helpful if we are trying to detect an excess signal (or narrow absorption feature) above (or below) a relatively flat continuum. The potential advantages of using an SDI calibration over an RDI one in this scenario would be that any time-varying systematic signals would be captured by the calibration, and there would not be any risk of introducing further noise due to spectral, airmass, or temporal differences between the calibrator and science target. Therefore, SDI would work well for the scenario there is a strong emission line or narrow absorption feature, such when searching for accretion signatures from protoplanets with high sensitivity.

One practical consideration when using adjacent wavelength bands for SDI is that, for a signal that is constant with wavelength, the companion would be at a fixed position relative to the star (and its PSF). Therefore, the phase at the $(u, v)$-sampling points will change slightly across the hyperspectral cube as the wavelength of the images changes with respect and the baseline lengths remain fixed. The magnitude of this effect is dependant on the width of the wavelength bins, which in the case of CHARIS' high-resolution K-band mode are small at ~22 nm. Therefore, one way to mitigate this issue is to only use adjacent bands when performing the SDI calibration. Then it can be assumed that there are minimal changes in the science signal from band to band.

Since a reference signal is being subtracted from a science signal, the efficacy of the calibration is proportional to the level of correlation between the two sets of kernel phases. Figure 8 shows the Pearson correlation coefficients for the time-averaged kernel phases of the (1) different bands of PSF calibrator, (2) different bands of the telluric, and (3) between the same band of the PSF and telluric calibrator. The binary is excluded since the kernel phases are dominated by the signal of HD 44927 B and not representative of the strength of any systematic noise. Strong correlations can be seen across the majority of the wavelength bands for both calibrators as well as between the same bands of the PSF and telluric. Band 17 shows a poor level of correlation in both calibrators (PSF: 0.02 and telluric: −0.35) and is even strongly anticorrelated in some of the bands of the telluric. Excluding band 17, the mean correlation for the PSF calibrator is 0.75 and for the telluric is 0.72. Both of these decrease to 0.68 and 0.59, respectively, with band 17 included, and it is currently not clear why the systematics in the final band are so different from the rest of the data. Though it may be due to the instrument (e.g., decreased throughput), or an artefact of the spectral extraction.

The mean intercalibrator value correlation coefficient is 0.78 (including band 17), higher than the intracalibrator values. However, if we consider only the nearest neighbors (the adjacent bands next to the target band), the mean correlation increases to 0.77 for both calibrators, which is comparable to the intercalibrator case. This suggests a simple SDI calibration strategy such that

$$C_{i,n} = K_{i,n} - 0.5 \times (K_{i,n-1} + K_{i,n+1}),$$ (6)

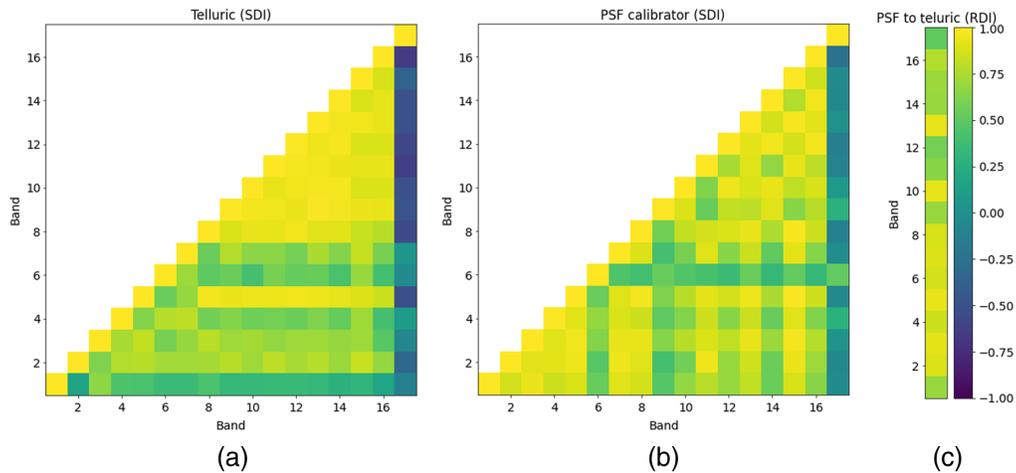

**Fig. 8** The Pearson correlation coefficient as measured for different bands of (a) the telluric calibrator data, (b) PSF calibrator, and (c) between the same band of the telluric and PSF. The coefficients were measured for the time-averaged kernel phase data. (a), (b) The strength of correlations between the different bands of the same dataset and give an indication of how well an SDI calibration may work. (c) The strength of correlation between the telluric and PSF calibrator, which is representative of the performance of the RDI calibration.

where $C$ is the calibrated kernel phase, $K$ is the raw kernel phase, $i$ is the kernel phase index, and $n$ is the band index. For the edges, the first and last band, this is modified to only use the available adjacent band. We apply this calibration to the time-averaged kernel phases, and the results are in Fig. 9, which shows the standard deviation of the time-averaged kernel phases computed across kernel phase index for the PSF and telluric, respectively, after calibration.

Both RDI and SDI calibrations reduce the standard deviations of the kernel phases (taken along the kernel phase index to estimate the level of systematic noise), indicating that the calibration is working for most bands. (We expect that the KPs for the PSF and telluric are dominated by systematic noise, since these set the mean KP values and random sources of error affect the scatter around the mean.) For the PSF star, SDI and RDI work equally well with the resulting average standard deviation (across the 17 wavelength bins) being almost identical at 18.48 deg and 18.32 deg. For the telluric star, this is not the case, and the SDI average standard deviation is significantly higher at 22.84 deg compared to 18.32 deg for RDI. Note that the average standard deviation in the RDI-calibrated PSF and telluric kernel phases are both the same, since one is simply the negative of the other. However, in the case of the telluric, most of the increase in noise comes from the very poor performance of the calibration in the first and last band. There, SDI increases the noise in the kernel phases compared to the raw data. Excluding the first and last bands, SDI has an average standard deviation of 17.7 deg compared to 15.6 deg for RDI.

Of the 17 bands, SDI outperforms RDI 8 times for the PSF calibrator and 9 times for the telluric calibrator. However, the calibrations give different results across the same bands of the two datasets. For example, while RDI performs better than SDI for both objects in band 13, SDI outperforms RDI for band 15 of the PSF calibrator but not the telluric. Fully understanding the variability of the two calibrations will be the subject of future work and requires a larger dataset.

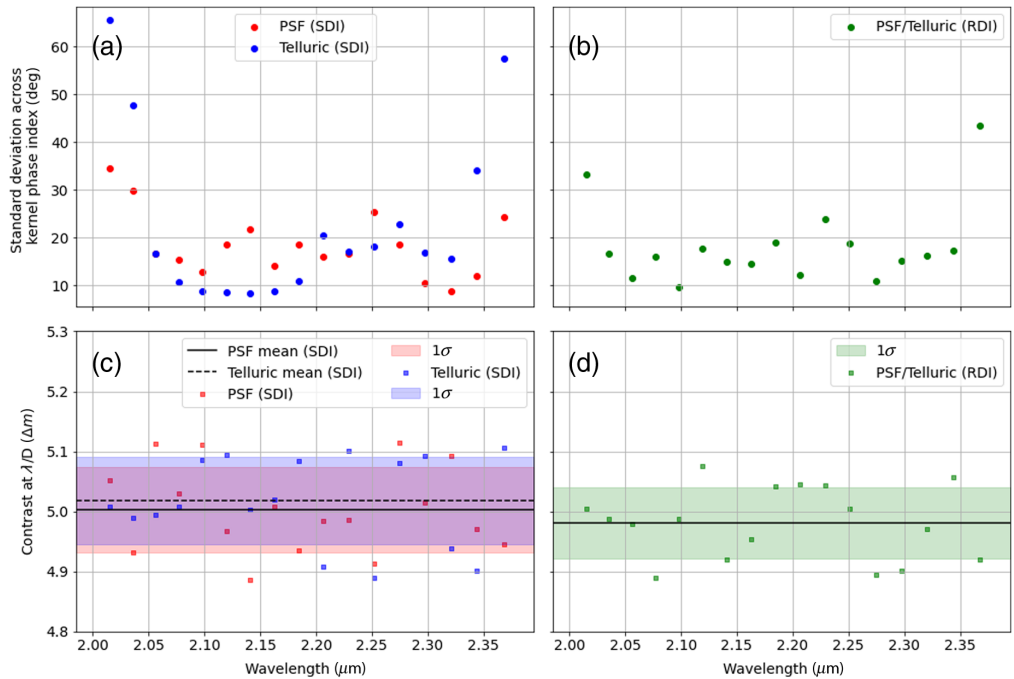

**Fig. 9** A comparison of the outcomes of RDI and SDI calibration of the PSF and telluric calibrator stars. (a), (b) Standard deviation across kernel phase index for the time-averaged kernel phases after calibration with each method. (c), (d) The $5 - \sigma$ contrast at $\lambda/D$ for each of the spectral bands computed using the chi-squared interval method. These contrast limits are calculated using the PSF and telluric calibrator datasets which consist of 42 and 52 frames, respectively. It can be seen that there is significant scatter in the quality of the final calibration as measured from the standard deviation. However, when converted into a contrast limit the difference is more modest. In practice, the contrast limits here represent an upper limit on the possible achievable contrast, since the chi-squared interval method assumes that the errors in the data are well understood.

However, the fact that they perform equally well on average suggests that it is possible to use SDI to calibrate kernel phases taken with an IFS, in the specific case where a sharp feature is expected in the spectrum of the science target. This could enable fewer dedicated reference PSF observations, allowing more time to build signal to noise on the science target. Future work will explore more sophisticated calibration strategies which leverage the full set of information in the IFS kernel phases to calibrate the data.

## 7 Simulating a CHARIS Kernel Phase Campaign

Simulating CHARIS KPI science observations requires a good model of the performance of the instrument. There are two key limitations to producing this.

1. The test data shown here cover only a short-time period, for bright targets observed in above average weather conditions, and so are not representative of the $S/N$ and AO performance that may be achieved for a typical night.
2. The degree to which systematic noise in the kernel phases is a limiting factor for performance is not well understood for other targets. This depends on the relative levels of photon, instrument and calibration errors present in the data, as well as observational constraints, such as the parallactic angle evolution and exposure time (which determine how many frames worth of data may be combined together for later analysis).

For (1), only further observations under a variety of conditions can inform our understanding of what contrasts can be reached, due to the difference in AO performance as a function of guide star brightness and observing conditions. The issue with (2) is somewhat mitigated by the parallactic angle evolution of the science target, since this will limit the maximum number of observations which can be time-averaged without blurring out circumstellar structure. For example, both the PSF calibrator and telluric star (which had 1.2 deg and 2.1 deg of parallactic angle evolution, respectively) were observed with 20.65 s exposure time and an approximate dead time of 5 s between exposures. This amounts to ~30 s per exposure or 120 exposures per minute.

Assuming that a planned, deeper science observation on a similar target would have 180 deg of parallactic angle evolution over the course of 6 h, this would mean ~4 exposures per 1 deg of angle change. Changes in parallactic angle are not linear over the course of the night, but under this simple assumption, binning to 5 deg would mean combining 20 frames (or 10 min of data). Binning further may risk averaging over a science signal that is rotating with the parallactic angle.

In practice, parallactic angle changes most rapidly close to transit, making this simple scenario unrealistic. For a real set of parallactic angles from a CHARIS half-night observation with 178 deg of evolution, near transit there are as few as four frames per 5 deg bin, while away from transit there can be as many as 90. Ultimately, the performance of CHARIS for each PA bin will be dictated by the relative contributions of random noise and systematics, with it being easier to reach the noise floor in snapshots far from transit, especially for faint targets.

In order to build a model of the noise properties of a typical set of observations, we start by considering the PSF calibrator. Figure 10 shows the standard deviation of the PSF calibrator kernel phases (band 8/2.162 nm) as a function of the number of combined frames. This standard deviation is calculated from the time-averaged kernel phases as a function of kernel phase index and is measured after binning the data. Both the uncalibrated data (green points) and the SDI data (orange points) are shown as well as a comparison to the final value for RDI (blue line).

In the presence of white noise, we would expect the standard deviation to decrease as the $\sqrt{n}$, where $n$ is the number of observations combined together. To assess this, we fit a simple model to the data:

$$\sigma_n = \frac{\sigma_0}{\sqrt{n}} + C, \tag{7}$$

with $\sigma_n$ denoting the standard deviation after binning $n$ frames, and $C$ denoting a fixed additive constant. The additive constant $C$ here represents residual systematic errors to which the standard deviation of the kernel phases will converge for a large number of observations. The model describes the binning behavior of the data well, indicating that some of the white noise is being binned over at least up to 22 data points per bin. The large disparity in the value of the constant between the SDI and uncalibrated data shows that the SDI (and RDI) calibrations are working to well, with ∼50% of the systematics removed by calibration. This can be seen by comparing the green and orange points in Fig. 10.

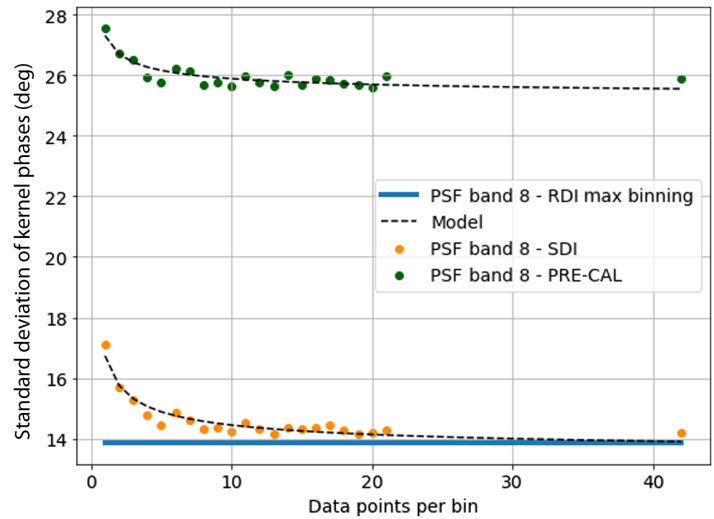

**Fig. 10** The standard deviation of the time-averaged kernel phases (calculated across the kernel phase indices) for band 8 of the PSF calibrator as a function of the number of averaged frames. This is shown for the uncalibrated data (green), KP-SDI (orange), and the final value for comparison using KP-RDI (dashed blue line). A model is fit to the orange points to predict the standard deviation (across kernel phase index) for larger numbers of combined frames, assuming Gaussian, white noise like behavior.

Using the model fit to the standard deviation of the SDI-calibrated PSF KPs, it is possible to estimate the maximum achievable contrast for a typical science observation assuming similar data quality. This implicitly assumes that the standard deviation of the KPs would converge similarly [the value $C$ in Eq. (7)] in the case of a longer sequence of observations as for the short sequence currently available. A further assumption is needed, discussed above, that typical observations consist of 180 deg parallactic angle evolution over 6 h and a binning of 5 deg in parallactic angle. Each binning has an independent noise realisation with a standard deviation given by Eq. (7). Figure 11 shows the resulting contrast (shaded regions) compared to the $5 - \sigma$ value measured from a single frame. These contrasts are computed by comparing the reduced chi-squared between the data and a null model (no signal), and the data and model binaries of different contrasts, PAs, and separations.

The sensitivities presented in Fig. 11 are relevant in the case of a search for line emission or a sharp absorption band feature, since any continuum emission would largely be removed by the SDI calibration (under the assumptions we make for the SDI calibration, using adjacent bands and small width of the wavelength bands, Sec. 6). Without any modeling of the noise, an RDI calibration is still required for signals with smoothly varying contrasts across wavelength (e.g., such as companions where you want to characterize the photosphere). In future work, it may be possible to model the noise properties of the data jointly with a science signal, using the wavelength information to separate the two signals. This may result in an "SDI-boost" to the sensitivity of the technique, while still preserving continuum information.

Nevertheless, when searching for Br-$\gamma$ line emission, the performance would greatly be improved by observing the target for 6 h (∼720 frames) compared to just one 20-s exposure, as expected. Both the minimum separation and the maximum contrast are improved, by over a magnitude for the latter. Several caveats are necessary; (i) this is only the case for data taken under similar conditions and with a similar target brightness (both in K-band for the science observations and R-band for the SCExAO wavefront sensor), (ii) the behavior of the noise is assumed to be white-noise limited up to 100 frames (extrapolating the model fit to the PSF band 8 data beyond the 42 frames which were available), and finally (iii) that the SDI calibration does not remove any of the science signal.

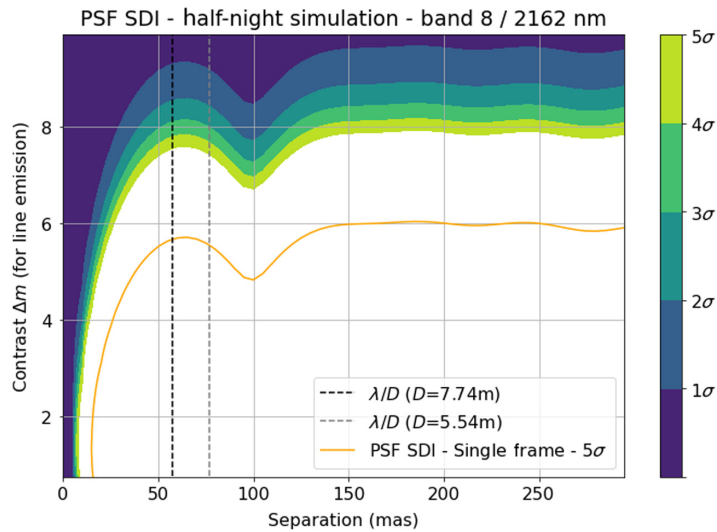

**Fig. 11** Estimated 5-$\sigma$ contrast for a single observation (orange line) and for a simulated half-night of observations (shaded region) consisting of a parallactic angle change of ~180 deg and 20 s exposure time. Dashed lines represent $\lambda/D$ for a maximum baseline of 5.8 m (dashed gray line) and 7.74 m (dashed black line), respectively. The KPs were calibrated using the SDI technique, which would remove any continuum emission present from the signal. The contrasts presented here are calculated with respect to the star and represent the sensitivity relative to a continuum signal from either line emission or a sharp absorption feature. Contrasts were computed by using the chi-squared interval method. The error values were taken from a model fit to the data from the PSF calibrator calibrated using the KP-SDI technique (Fig. 10). By making use of this fit, it is implicitly assumed that the noise properties of the short sequence of CHARIS data presented here are representative of those for a longer observation. Increasing the observation time from a single frame (20-s exposure) to a half-night (6 h; 21,600 s) increases the achievable contrast by over a magnitude for separations above 25 mas. Below 25 mas, the closest detectable separation is greatly improved, increasing the sensitivity to close-in companions. For the purposes of future observations, these contrast levels represent an upper limit on potential performance under similar conditions, since the chi-squared interval method assumes good knowledge of the error bars on the data and that they are Gaussian.

While (iii) may be an issue when recovering continuum emission, this simulation is broadly applicable to a scenario where the underlying signal is an excess above continuum emission in a single band. In the case of a Br-$\gamma$ search, this is expected to work well due to the relatively small size of the spectral bands (~22 nm) for CHARIS' high-resolution K-band mode. Future work will study how a full calibration can be conducted without losing continuum information using a more sophisticated KP-SDI scheme. In practice, this simulation may be considered an expected upper limit on performance, both due to the assumptions above and the dependence of the contrast curve generation method on reliable error bar estimation.

## 8 Conclusion

In this study, we conducted KPI observations using the CHARIS/SCExAO IFS instrument, by recovering the low-contrast binary HD 44927 AB and generating a spectrum of B in contrast space. This is the first demonstration of simultaneous detection and characterization using IFS data with super-resolution provided by KPI. Spectrally dispersed kernel phase opens up the use of the technique for future spectroscopic observations at angular resolutions beyond the limits of conventional data analysis techniques. Additionally, we have demonstrated that via the use of an IFS, an SDI calibration can be performed for kernel phase. Currently, this calibration is only applicable for the science case of a strong differential signal, such as line emission or a sharp absorption feature, localized over a narrow wavelength range. This is because the current implementation removes any continuum signal which may be present in the data. In future work, simultaneous modeling of the kernel phase errors and science signal may provide avenues to apply KP-SDI to a broader range of science cases.

A key example of where KP-SDI may currently be helpful is the search for accretion-tracing line-emission around young stars. Direct evidence of accretion from the disk onto young planetary-mass objects is critical to answering many of the outstanding questions of planet formation. Such "caught-in-formation" systems are exceptionally rare, with only a handful of examples verified. Further sensitive searches using KP-SDI may yield new candidate companions with smaller orbital separations than are accessible with conventional direct-imaging searches. The resolution advantage of KPI would bring us closer to the regime where core accretion is thought to be efficient, allowing for a better understanding of its detailed physics processes.

## 9 Appendix A

Figure 12 shows the full corner plot for the MCMC fit of a companion model to the RDI calibrated kernel phases for the HD 44927 observations. In total nineteen parameters are fit: a common separation and position angle followed by a contrast for each of the seventeen wavelength bins. The Gaussian posteriors indicate a good fit and little degeneracy between the parameters.

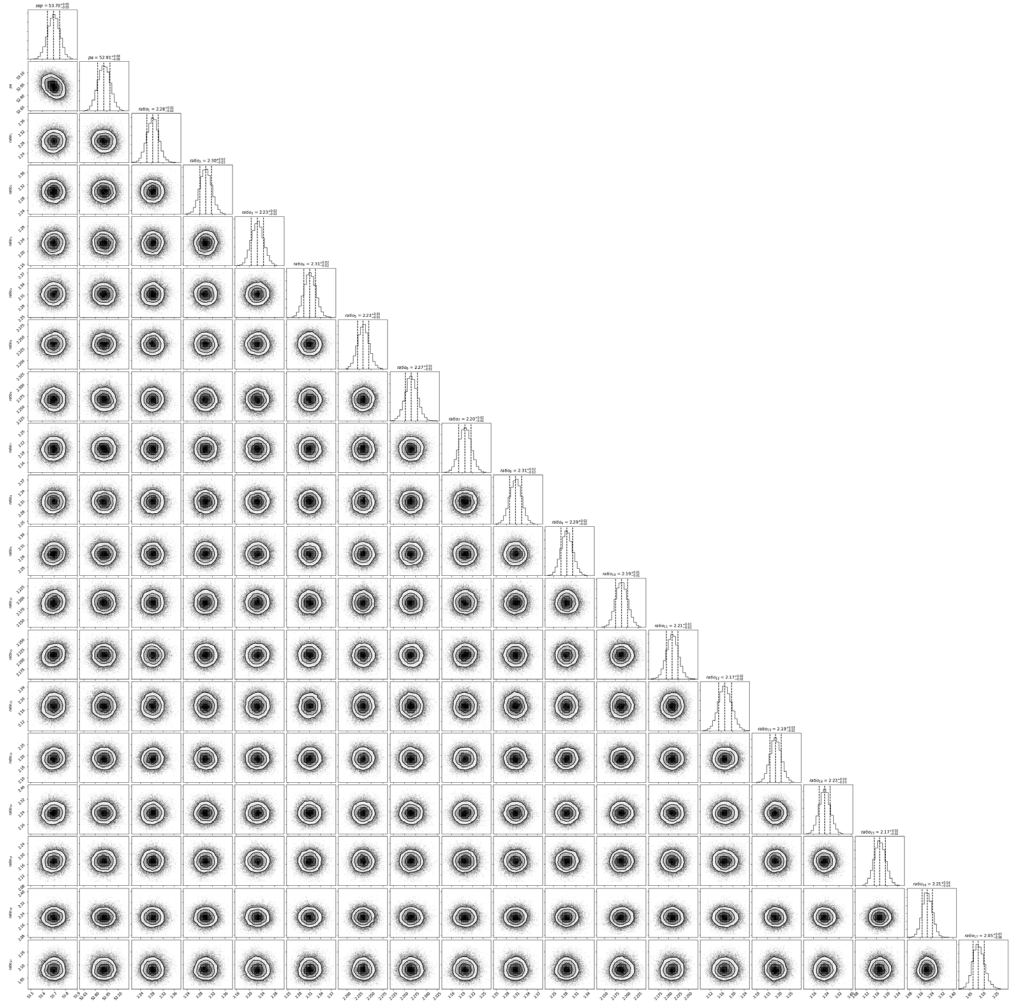

**Fig. 12** Corner plot for the MCMC fit of the binary HD 44927. The first two parameters are the separation and PA, followed by the flux ratio (primary/secondary) for each of the 17 spectral bands comprising the high-res K-band mode. The flux ratios are arranged in order of increasing wavelength from left to right.

## Acknowledgments

This research was funded by Heising-Simons Foundation (Grant No. 2020-1825). Based on data collected at Subaru Telescope, which is operated by the National Astronomical Observatory of Japan. The development of SCExAO is supported by the Japan Society for the Promotion of Science (Grant Nos. 23340051, 26220704, 23103002, 19H00703, 19H00695, and 21H04998), the Subaru Telescope, the National Astronomical Observatory of Japan, the Astrobiology Center of the National Institutes of Natural Sciences, Japan, the Mt. Cuba Foundation, and the Heising-Simons Foundation. The authors would wish to recognize and acknowledge the very significant cultural role and reverence that the summit of Maunakea has always had within the indigenous Hawaiian community and are most fortunate to have the opportunity to conduct observations from this mountain. This research has made use of the Washington Double Star Catalog maintained at the U.S. Naval Observatory. The authors would like to thank the anonymous reviewers for their valuable comments which have greatly improved the paper. The authors have no relevant financial interests in this manuscript and no other potential conflicts of interest to disclose.

## Code, Data, and Materials Availability

The data from this study are available on reasonable request to the authors.

**Alexander Chaushev** is a postdoctoral scholar at the University of California Irvine.

Biographies of the other authors are not available.